\documentclass[preprint,12pt]{elsarticle}

\usepackage{amssymb}
\usepackage{amsthm}

\usepackage{float}
\usepackage{color}
\usepackage{amsmath}
\usepackage{graphicx, subfigure}
\usepackage{pslatex}
\usepackage{relsize}
\usepackage{bm}
\usepackage{verbatim}

\usepackage{hyperref}

\usepackage{booktabs}
\usepackage{xcolor}
\usepackage{soul}
\usepackage{ulem}

\usepackage[utf8]{inputenc}
\usepackage[T1]{fontenc}

\biboptions{numbers,sort&compress}

\begin{document}

\begin{frontmatter}



\title{VelCrys: Interactive web-based application to compute acoustic wave velocity in crystals and its magnetic corrections}


\author[a]{P.~Nieves\corref{author}}
\author[b]{I. Korniienko}
\author[c]{A. Fraile}
\author[a]{J.M. Fernández}
\author[a]{R. Iglesias}
\author[b]{D. Legut}

\cortext[author] {Corresponding author.\\\textit{E-mail address:} nievespablo@uniovi.es}
\address[a]{ Departamento de F\'{i}sica, Universidad de Oviedo, C. Leopoldo Calvo Sotelo, 18, 33007, Oviedo, Spain}

\address[b]{IT4Innovations, V\v{S}B - Technical University of Ostrava, 17. listopadu 2172/15, 70800 Ostrava, Czech Republic}

\address[c]{ Material Physics Center (MPC) / Centro de Física de Materiales(CFM) CSIC-UPV/EHU, San Sebastián, Spain}


\begin{abstract}

We present VelCrys, a web-based interactive tool, that allows to perform further post-processing of the elastic tensor in order to compute and plot the group velocity of the acoustic waves for any crystal symmetry. We also implemented the calculation of  effective magnetic corrections to the elastic tensor and corresponding fractional change in group velocity under a magnetic field. We apply it to dry sandstone, cubic CoPt and hcp Co to show some of the program features. In the analysis of magnetic corrections, we find complex landscapes of fractional change in group velocity as a function of ray direction, as well as a field dependence consistent with Simon effect.

\end{abstract}

\begin{keyword}
Acoustic \sep Crystals \sep Magnetism

\end{keyword}

\end{frontmatter}


\section{Introduction}
\label{section:Intro}

The characterization of wave velocity is particularly useful in many fields like  geophysics\cite{wang2023,REGAN1984227,CARA1987246,thomsen,Schimmel}, communication technology\cite{Yang_2022} or biosensors\cite{HUANG2021100041}. The way acoustic waves propagate in crystals can be highly anisotropic, and thus it is different and more complicated than in isotropic media \cite{Landau}. Providing the phase velocity direction ($\boldsymbol{k}$), one finds three possible modes for an acoustic wave: one quasi-compressional wave (qP) and two quasi-shear waves (qS1 and qS2)\cite{wang2023}. Over the last five decades, great efforts have been made to develop theoretical techniques to compute wave velocity for a general anisotropic media\cite{Neigh1967,Cerveny1972,CRAMPIN1981343,Bing_Zhou_2004,wang2023}. For instance, Zhou \textit{et al.}\cite{Bing_Zhou_2004} proposed the eigenvalue and  eigenvector methods to calculate the group velocity. A more manageable approach has been recently proposed by Wang \textit{et al.}\cite{wang2023} based on analytical formulas for group velocity, which can deal with the singular points and multivalue problems of the qS wave. 

Magnetic effects on wave velocity in magnetostrictive materials are interesting from a fundamental and applied point of view\cite{Rouchy1979,ROUCHY198069,booktremolet,DelMoral2008}. The effect of a magnetic field on wave velocity was originally discovered by Simon in 1958 (Simon effect)\cite{Simon+1958+84+89}. Additionally, other magnetic effects have been identified like isotropic exchange effects\cite{nieves_sound2022}, anisotropic morphic effects\cite{Mason1951b,Eastman,Rouchy1979}, rotational-magnetostrictive effects \cite{Rouchy1979,ROUCHY198069}, and $\Delta s$-effect \cite{ROUCHY198069}, which have been experimentally observed in  ultrasonic elastic waves\cite{ROUCHY198069,DUTREMOLETDELACHEISSERIE198277,ROUCHY198159,sakurai,ALERS195921,Dietz1976,booktremolet,kingner,SEAVEY1972219,DelMoral2008}.  In 1979, a comprehensive theoretical description of these effects were reported by Rouchy \textit{et al.} for cubic crystals \cite{Rouchy1979}. Later, Rinaldi and Turilli showed that linear magnetic effects on wave velocity can  be equivalently taken into account as effective corrections to the elastic constants \cite{rinaldi1985}. The addition of these corrections to the elastic tensor combined with Wang \textit{et al.} method \cite{wang2023} to compute group velocity may provide a practical way to efficiently compute the fractional change in group velocity due to linear magnetic effects for arbitrary wave propagation direction and applied magnetic field. Hence, it may facilitate the analysis of this phenomenon beyond the cases with wave propagation and magnetic field directions along  high symmetry crystallographic directions studied so far, as well as crystals with symmetry lower than cubic \cite{Rouchy1979,ROUCHY198159,nieves_sound2022,booktremolet}. Such kind of approach may be exploited in novel magneto-acoustic phenomena like acoustic spin pumping \cite{Uchida2011}, magnetization switching induced by sound waves \cite{Camara,Kovalenko,Thevenard2013,Thevenard2016,Vlasov,Stupakiewicz2021,Weiyang,strungaru2023route} and  picosecond ultrasonics \cite{MATTERN2023100503}.

The development advanced visualization tools in databases can help to accelerate the design of new materials in different areas of condensed matter physics\cite{Mat_Proj_1,CURTAROLO2012218,Scheidgen2023}. For example,   ELATE is an online user-friendly interface that has been implemented to analyze the elastic 
tensor and visualize anisotropic mechanical properties \cite{elate}. Similarly, MAELASviewer was developed to visualize anisotropic magnetostrictive properties\cite{maelasviewer,maelasviewerGithub,maelasviewerWeb}. In this work, we present VelCrys a web-based interactive tool that performs further post-processing of the elastic tensor in order to compute and plot the sound velocity of the three kinds of acoustic waves (qP, qS1, and qS2) for any crystal symmetry based on the analytical formulas derived by Wang \textit{et al.}\cite{wang2023}. Additionally, if the magnetoelastic constants are also provided, then it can calculate the effective magnetic corrections to the elastic tensor and corresponding fractional change in sound velocity under a magnetic field using Rinaldi and Turilli approach\cite{rinaldi1985}.

\section{Methodology}
\label{section:method}

\subsection{Software details}
\label{software}

VelCrys is created with Dash which is an efficient Python framework for building web applications \cite{dash}. Dash uses Flask as the web framework \cite{flask}. Visualization of 3D surfaces is done using the Python open source graphing library Plotly \cite{plotly}. We also make use of NumPy library for some mathematical operations.  The online application is currently available at \url{http://www.md-esg.eu/velcrys/}, while the open source Python module is available at GitHub repository \cite{VelCrys}.
 VelCrys can be also used offline by executing the Python module\cite{VelCrys}.

\subsection{Calculation of sound velocity and magnetic corrections}
\label{theory}

The calculation of group velocity $v$ for any crystal symmetry is performed by computing the analytical solution derived by Wang \textit{et al.}\cite{wang2023} which makes use of the elastic tensor $C_{ij}$ as input parameter
\begin{equation}
    v_l(C_{ij})= \frac{\partial c}{\partial n_l} =\frac{\partial\omega}{\partial k_l},\quad l=x,y,z
    \label{eq:dvg}
\end{equation}
where $c=\omega/k$ is the phase velocity, and $\boldsymbol{n}=\boldsymbol{k}/k$ is the unit vector of phase velocity direction (ray direction).

The effective magnetic corrections ($\Delta C_{ij}$) to elastic tensor  are obtained through the formulas derived by Rinaldi and Turilli which have the form \cite{rinaldi1985}
\begin{equation}
    \Delta C_{ij} = \frac{1}{M_s^2}b_{pqi}b_{rsj}\alpha_p\alpha_r\chi_{qs},
    \label{eq:dC}
\end{equation}
where $b$ are magnetoelastic constants, $M_s$ is saturation magnetization, $\alpha$ is direction cosine of magnetization at equilibrium, and $\chi$ is the susceptibility tensor, see Ref.\cite{rinaldi1985} for details about subscript indexes. The corresponding fractional change in group velocity due to magnetic effects $(v-v_0)/v_0$ is obtained by computing Wang \textit{et al.} formulas\cite{wang2023} using the elastic tensor including effective magnetic corrections ($C_{ij}+\Delta C_{ij}$), that leads to $v$, and elastic tensor without effective magnetic corrections ($C_{ij}$) which gives $v_0$, that is
\begin{equation}
    \frac{v-v_0}{v_0}=\frac{v(C_{ij}+\Delta C_{ij})-v(C_{ij})}{v(C_{ij})}.
    \label{eq:dv}
\end{equation}
This option is currently supported only for Cubic I (point groups $432$, $\bar{4}3m$, $m\bar{3}m$, space group numbers from 207 up to 230), and Hexagonal I (point groups $6mm$, $622$, $\bar{6}2m$, $6/mmm$, space group numbers from 177 up to 194) symmetries\cite{rinaldi1985}. Note also that Rinaldi and Turilli expressions does not include higher order corrections like  morphic and rotational-magnetostrictive effects \cite{nieves_sound2022}. These limitations are discussed in Section \ref{section:CoPt}.

\section{Results}
\label{section:results}

In this section, we perform some tests of the methodology implemented in VelCrys.

\subsection{Dry sandstone}
\label{section:sandstone}

\begin{figure}[h!]
\centering
\includegraphics[width=0.7\columnwidth ,angle=0]{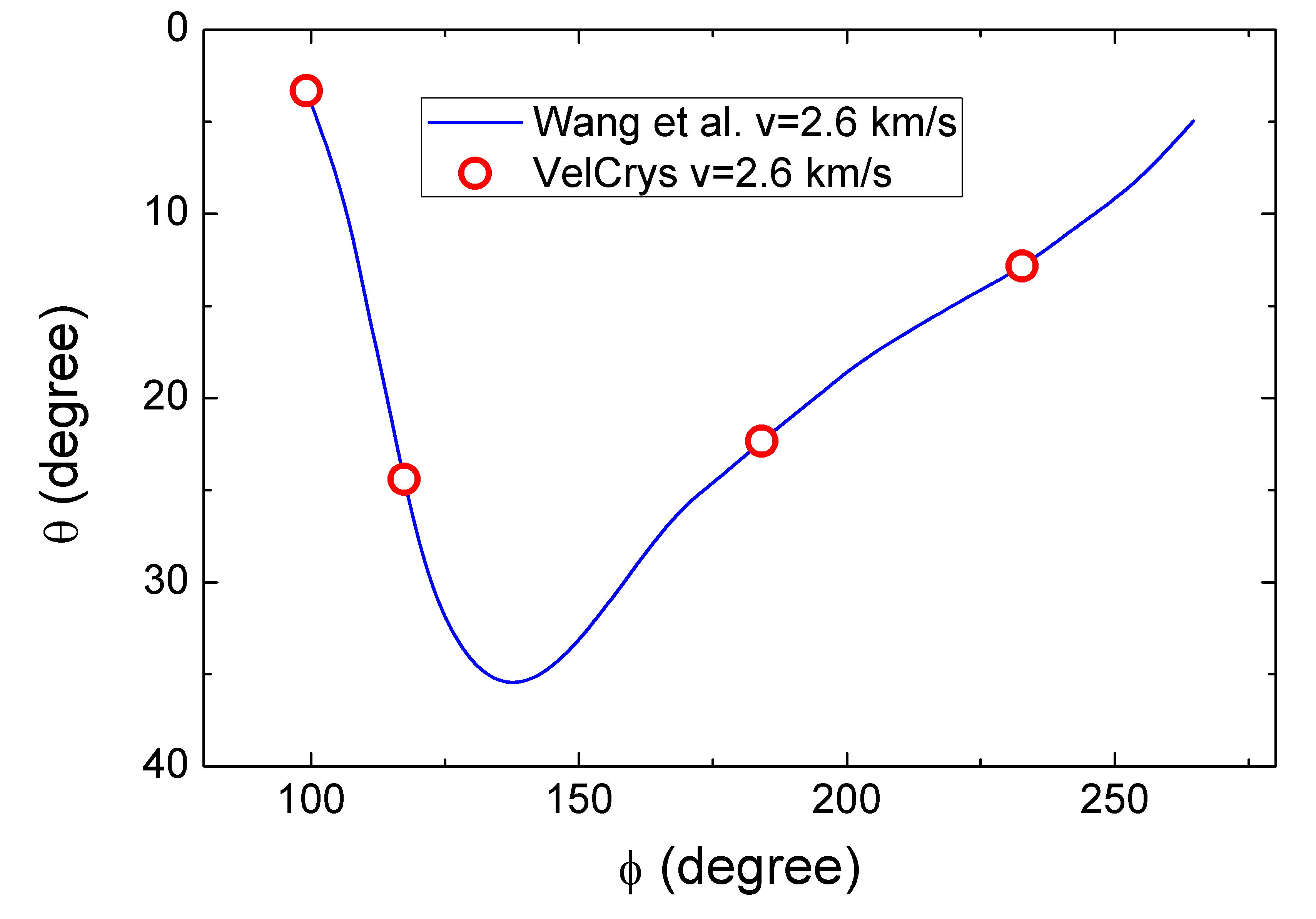}
\caption{Isoline of the group velocity ($v=2.6$ km/s) of the qP wave for dry sandstone with varied ray angle ($\phi,\theta$) obtained with VelCrys (red open circles) and by Wang \textit{et al.}\cite{wang2023} (blue solid line). }
\label{fig:sandstone}
\end{figure}

\begin{figure}[h!]
\centering
\includegraphics[width=0.5\columnwidth ,angle=0]{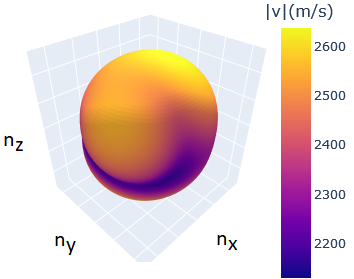}
\caption{Group velocity landscape as a function of the ray direction (phase velocity direction) $\boldsymbol{n}=\boldsymbol{k}/k=(\sin\theta\cos\phi,\sin\theta\sin\phi,\cos\theta)$ of the qP wave for dry sandstone generated with VelCrys. The color of the surface corresponds to the magnitude of group velocity $\vert v\vert$.}
\label{fig:sandstone3D}
\end{figure}

Sandstone is sedimentary rock widely distributed in planet Earth mostly composed of sand-sized minerals or rock grains
 cemented by siliceous, argillaceous, and/or calcareous materials \cite{Tian2016}. Its mechanical properties depend highly
upon the degree of cementation and grain composition \cite{Tian2016} which are affected by temperature and burial history \cite{Tian2016,QIN}. In this first example, we verify the correct implementation in VelCrys of Wang \textit{et al.}\cite{wang2023,Wang_code} methodology by computing the group velocity of complex dry sandstone and compare with their  results\cite{wang2023}. The elastic constants of this material are shown in Table \ref{table:data_properties_sandstone}, while the mass density is $\rho=2080$ kg/m$^3$ at room temperature\cite{wang2023}. In Fig.\ref{fig:sandstone}, we compare the group velocity for a qP wave obtained with VelCrys and reported by Wang \textit{et al.}\cite{wang2023}, finding good agreement. A 3D plot generated with VelCrys is shown in Fig.\ref{fig:sandstone3D}, where one can see the group velocity landscape of the qP wave as a function of the ray direction (phase velocity direction), again consistent with Ref. \cite{wang2023}.

\begin{table*}[ht]
\caption{Elastic constants ($C_{ij}=C_{ji}$) of dry sandstone \cite{wang2023}.}
\label{table:data_properties_sandstone}
\centering
\resizebox{\textwidth}{!}{
\begin{tabular}{cc|cc|cc|cc|cc|cc}
\toprule
$C_{ij}$ &  (GPa) &
$C_{ij}$ &  (GPa) & 
$C_{ij}$ &  (GPa) &
$C_{ij}$ &  (GPa) &
$C_{ij}$ &  (GPa) &
$C_{ij}$ &  (GPa) \\

\midrule
 $C_{11}$  & 10.3 &  $C_{12}$  & 0.9 & $C_{13}$  & 1.3 & $C_{14}$  & 1.4 & $C_{15}$  & 1.1 & $C_{16}$  & 0.8 \\
  &  &  $C_{22}$  & 10.6 & $C_{23}$  & 2.1 & $C_{24}$  & 0.2 & $C_{25}$  & -0.2 & $C_{26}$  & -0.6 \\
   &  &    &  & $C_{33}$  & 14.1 & $C_{34}$  & 0.0 & $C_{35}$  & -0.5 & $C_{36}$  & -1.0 \\
   &  &    &  &   &  & $C_{44}$  & 5.1 & $C_{45}$  & 0.0 & $C_{46}$  & 0.2 \\
     &  &    &  &   &  &  &  & $C_{55}$  & 6.0 & $C_{56}$  & 0.0 \\
        &  &    &  &   &  &  &  &  &  & $C_{66}$  & 4.9 \\
\bottomrule
\end{tabular}
}
\end{table*}

\begin{table*}[ht]
\caption{Experimental elastic constants ($C_{ij}$), anisotropic magnetoelastic constants ($b$),  anisotropic magnetostrictive coefficients ($\lambda$), magnetocrystalline anisotropy constant ($K_1$), saturation magnetization ($M_s$) and mass density ($\rho$) for cubic single crystals of disordered equiatomic Co-Pt alloy at room temperature \cite{ROUCHY198069,CoPt_exp}.}
\label{table:data_properties_CoPt}
\centering
\resizebox{\textwidth}{!}{
\begin{tabular}{cc|cc|cc|c|c|c}
\toprule
 $C_{ij}$
&  (GPa)  &
$b$	& (MPa)  &
$\lambda$	& ($\times10^{-6}$) 	& $K_1$(kJ/m$^3$) &  	 $\mu_0M_s$(T)  &  $\rho$(kg/m$^3$)\\

\midrule
 $C_{11}$  & 289.7  & $b_1$ &  -35 &$\lambda_{100}$ & 210  &  -40 &   0.942 &	16696.7 \\
                   $C_{12}$ & 178.5 & $b_2$ &  11.9 & $\lambda_{111}$ & -53    &  &   \\
			$C_{44}$  & 124.1 & & & & & &  &  \\

\bottomrule
\end{tabular}
}
\end{table*}

\begin{table*}[ht]
\caption{Theoretical expression for each magnetic contribution to fractional change in group velocity $(v-v_0)/v_0$ in cubic crystal \cite{Rouchy1979,ROUCHY198069,nieves_sound2022}. Here, $\boldsymbol{u}$ corresponds to the displacement vector, while other quantities are defined in the main text. }
\label{table:theory}
\centering
\resizebox{\textwidth}{!}{
\begin{tabular}{ccc|c|c|c|c|c}
\toprule
 $\boldsymbol{k}\parallel$
&  $\boldsymbol{u}\parallel$  &
$\boldsymbol{H}\parallel$	& $G(m)$  &
$R(\lambda)$ & $S(H)$ 	& $(l-l_0)/l_0$ & Wave\\

\midrule
 $[001]$  & $[001]$  & $[001]$ &  $\frac{2 m_1^{\gamma,2}}{3C_{11}}$ & 0 & 0  &  $\lambda_{100}$ &  qP  \\
$[001]$  & $[001]$  & $[100]$ &  $-\frac{ m_1^{\gamma,2}}{3C_{11}}$ & 0 & 0  &  $-\frac{\lambda_{100}}{2}$ &  qP  \\
 $[001]$  & $[100]$  & $[001]$ &  $-\frac{m_3^{\gamma,2}}{12C_{44}}$ & $\frac{3}{2}(\lambda_{111}-\lambda_{100}) $ & $- \frac{(b_2)^2}{2C_{44}\mu_0M_s\left(H+H_D+\frac{2K_1}{\mu_0M_s}\right)}$  &  $\lambda_{100}$ &  qS1  \\
 $[001]$  & $[100]$  & $[100]$ &  $-\frac{m_3^{\gamma,2}}{12C_{44}}$ & $\frac{3}{2}(\lambda_{100}-\lambda_{111}) $ & $- \frac{(b_2)^2}{2C_{44}\mu_0M_s\left(H+M_s+H_D+\frac{2K_1}{\mu_0M_s}\right)}$  &  $-\frac{\lambda_{100}}{2}$ &  qS1  \\

\bottomrule
\end{tabular}
}
\end{table*}

\begin{figure}[h!]
\centering
\includegraphics[width=0.7\columnwidth ,angle=0]{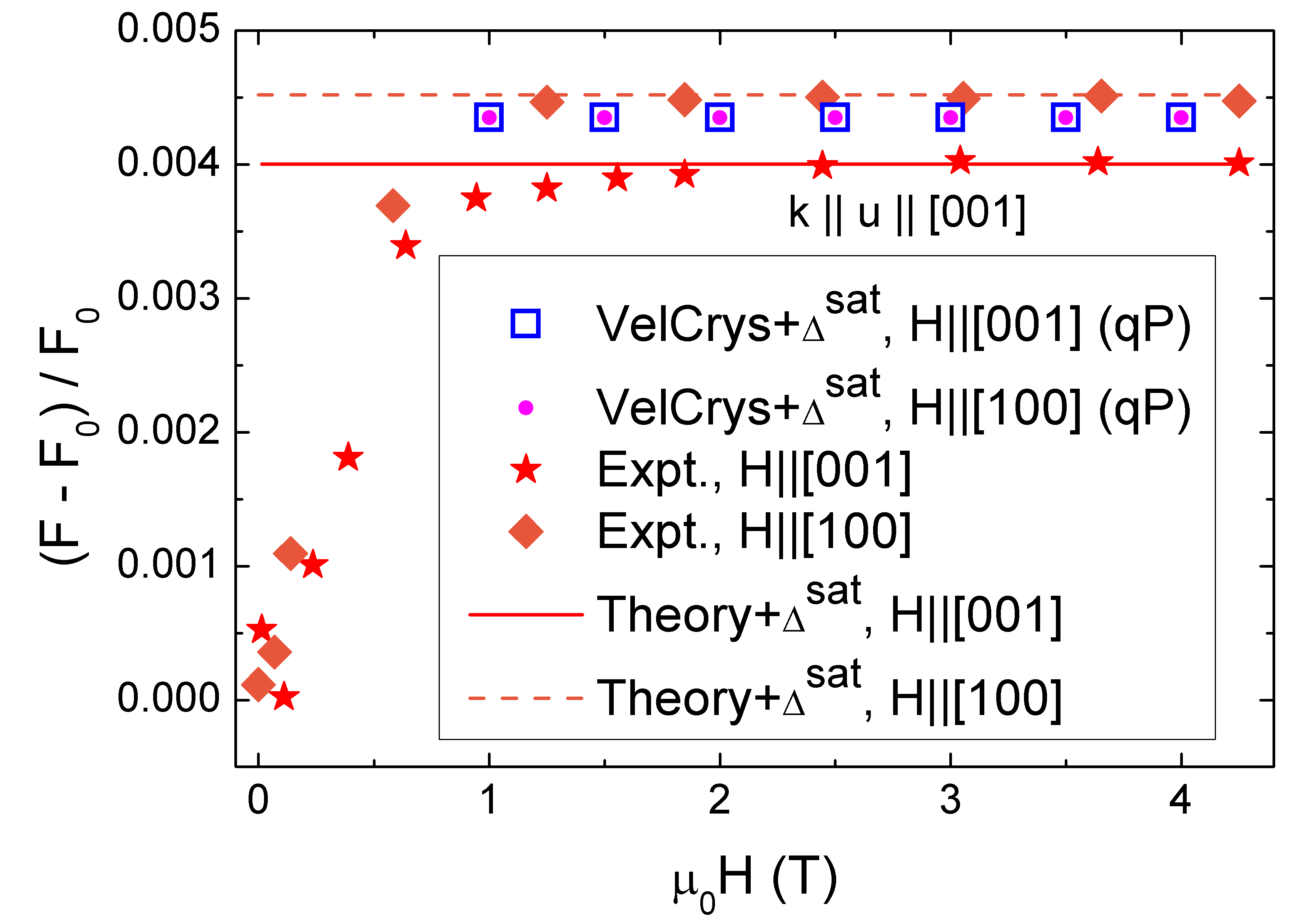}
\caption{Fractional change in ultrasound frequency for a longitudinal wave (qP) under an applied magnetic field parallel or perpendicular to the wave propagation vector for cubic single crystals of disordered Co-Pt alloy. Red stars and orange diamonds show experimental data from Ref.\cite{ROUCHY198069} for the cases $H\parallel[001]$ and $H\parallel[100]$, respectively. Red solid and orange dashed lines stand for the theoretical values obtained from Eq.\ref{eq:dF} and Table \ref{table:theory}. Open blue squares and solid pink circles give the values derived from the computed fractional change in group velocity $(v-v_0)/v_0$ by VelCrys via Eq.\ref{eq:dv}. We added experimentally observed $\Delta s$-effect at saturated state ($\Delta^{sat}=0.00435$) both in the theoretical values and VelCrys data.}
\label{fig:CoPt_long}
\end{figure}

\begin{figure}[h!]
\centering
\includegraphics[width=0.7\columnwidth ,angle=0]{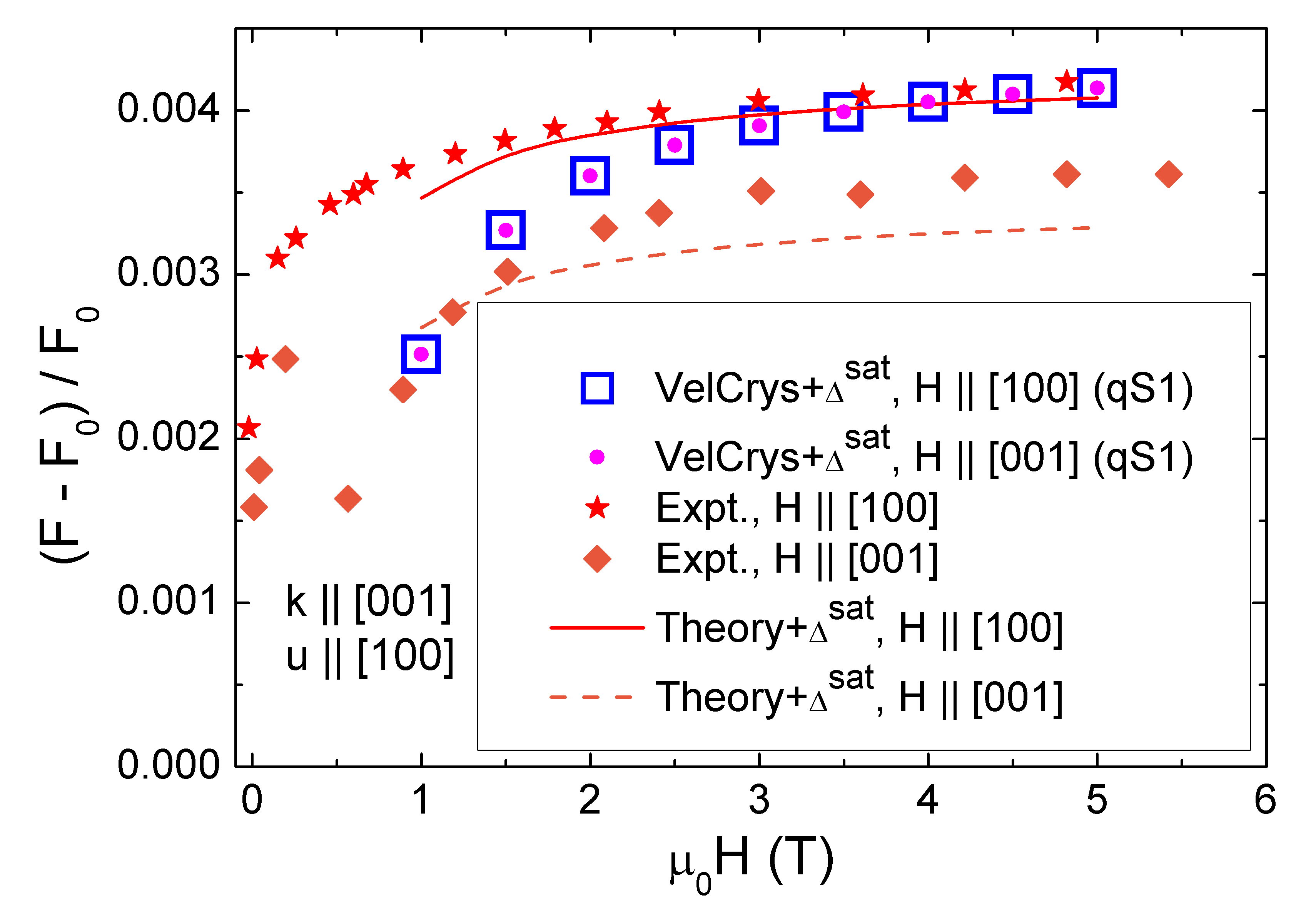}
\caption{Fractional change in ultrasound frequency for a transverse wave (qS1) under an applied magnetic field parallel or perpendicular to the wave propagation vector for cubic single crystals of disordered equiatomic Co-Pt alloy. Red stars and orange diamonds show experimental data from Ref.\cite{ROUCHY198069} for the cases $H\parallel[100]$ and $H\parallel[001]$, respectively. Red solid and orange dashed lines stand for the theoretical values obtained from Eq.\ref{eq:dF} and Table \ref{table:theory}. Open blue squares and solid pink circles give the values derived from the computed fractional change in group velocity $(v-v_0)/v_0$ by VelCrys via Eq.\ref{eq:dv}.  We added experimentally observed $\Delta s$-effect at saturated state ($\Delta^{sat}=0.00435$) both in the theoretical values and VelCrys data.}
\label{fig:CoPt_trans}
\end{figure}

\begin{figure}[h!]
\centering
\includegraphics[width=\columnwidth ,angle=0]{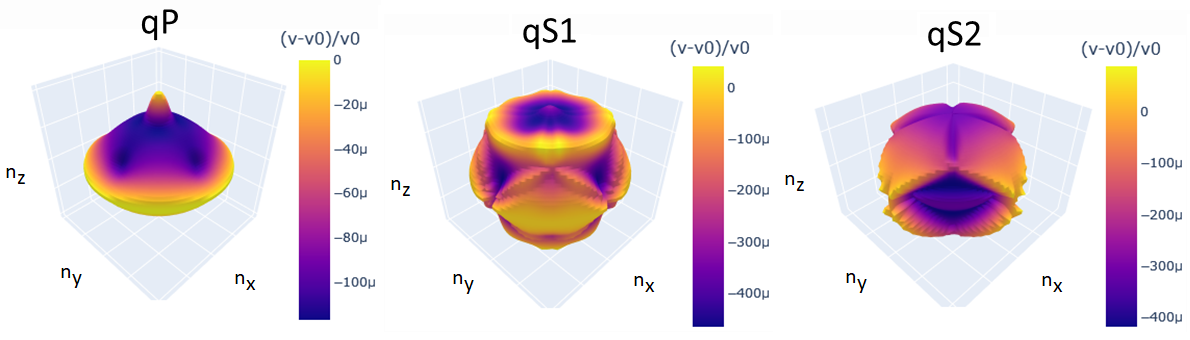}
\caption{Landscape of the fractional change in group velocity $(v-v_0)/v_0$ for the 3 types of waves (qP, qS1 and qS2) as a function of the ray direction (phase velocity direction) with applied magnetic field $\mu_0\boldsymbol{H}=(0,0,5)$ T for cubic single crystals of disordered equiatomic Co-Pt alloy generated with VelCrys. }
\label{fig:CoPt_3d}
\end{figure}

\subsection{Co-Pt alloy}
\label{section:CoPt}

Co-Pt (and Fe-Pt) are very interesting alloys since supported nanostructures of these alloys present direct link between atom arrangements and magnetic behavior. Additionally, both alloys are today model systems in the field of nanoalloys, due to the diversity of atom arrangements present either in bulk state
or specific to the nanoscale (disordered fcc structures, chemically ordered L1$_0$, L1$_2$, core-shell, five-fold structures - icosahedral or decahedral, etc). Importantly, Co-Pt alloys that can be used as permanent-magnet materials are found around the equiatomic composition where the disordered A1 (fcc) phase transforms into the ordered L1$_0$ phase \cite{darling}. Here, we further test VelCrys capabilities by evaluating magnetic corrections to group velocity for cubic single crystals of disordered equiatomic Co-Pt alloy, and comparing the results with theory and experiment \cite{ROUCHY198069,CoPt_exp}. The fractional change in group velocity $(v-v_0)/v_0$ due to magnetic effects may be written as \cite{Rouchy1979,nieves_sound2022}
\begin{equation}
 \frac{v-v_0}{v_0} = \frac{F-F_0}{F_0} +\frac{l-l_0}{l_0},
\label{eq:dv}
\end{equation}
where $F$ is the frequency of sound wave measured in ultrasound experiments (for instance, by the pulse echo
method\cite{ROUCHY198069,ROUCHY198159}) and $l$ is the length in the direction of wave propagation. The fractional change of frequency could be further splitted into \cite{Rouchy1979,nieves_sound2022,booktremolet,DelMoral2008,ROUCHY198069}
\begin{equation}
 \frac{F-F_0}{F_0} = G(m) + R(\lambda) + S(H) + \Delta,
\label{eq:dF}
\end{equation}
where $G(m)$ accounts for higher order correction arising from the morphic coefficients $m$, $R(\lambda)$ comes from  rotational and magnetostrictive effects, $S(H)$ describes the Simon effect that depends on the external magnetic field $\boldsymbol{H}$ \cite{Simon+1958+84+89}, and $\Delta$ corresponds to other possible effects like the $\Delta s$-effect associated to domain walls displacement in the low magnetic field regime \cite{ROUCHY198069}. The anisotropic part of the fractional change in length reads \cite{nieves_sound2022,booktremolet}
\begin{equation}
\begin{aligned}
     \frac{l-l_0}{l_0}\Bigg\vert_{\boldsymbol{\beta}}^{\boldsymbol{\alpha}} 
     & =\lambda^{\gamma,2}\left(\alpha_x^2\beta_{x}^2+\alpha_y^2\beta_{y}^2+\alpha_z^2\beta_{z}^2-\frac{1}{3}\right)\\
     & + 2\lambda^{\epsilon,2}(\alpha_x\alpha_y\beta_{x}\beta_{y}+\alpha_y\alpha_z\beta_{y}\beta_{z}+\alpha_x\alpha_z\beta_{x}\beta_{z}),
    \label{eq:delta_l_cub_I}
\end{aligned}
\end{equation}
where $\boldsymbol{\alpha}$ is the direction cosine of magnetization, $\boldsymbol{\beta}$ is the normalized vector along the measuring length direction, and $\lambda^{\gamma,2}=3\lambda_{100}/2$ and $\lambda^{\epsilon,2}=3\lambda_{111}/2$ are the magnetostrictive coefficients. In this example, we consider longitudinal (qP) and transverse (qS) elastic waves propagating along the crystallographic direction $[001]$ ($\boldsymbol{k}\parallel[001]$), where the magnetic field is  applied either along $\boldsymbol{H}\parallel[001]$ or $\boldsymbol{H}\parallel[100]$. In Table \ref{table:theory}, we present the theoretical expression for each magnetic contribution to fractional change in group velocity $(v-v_0)/v_0$ in cubic crystal \cite{Rouchy1979,ROUCHY198069,nieves_sound2022} for these cases. A comparison between experiment, theory and VelCrys is shown in Figs.\ref{fig:CoPt_long} and \ref{fig:CoPt_trans}, finding consistent results. In function $G(m)$, we use experimental values of the morphic coefficients $m_1^{\gamma,2}=-0.15$ GPa, and $m_3^{\gamma,2}=0.77$ GPa, while for the other properties we use the experimental values shown in Table \ref{table:data_properties_CoPt}\cite{ROUCHY198069}. The fractional change in frequency $F$ is indirectly obtained in VelCrys by using the calculated fractional change in group velocity in Eq.\ref{eq:dv}.  Here, the experimental observed $\Delta s$-effect, about $\Delta=0.00435$ at saturated state\cite{ROUCHY198069}, is included both in theory and VelCrys. For the sake of simplicity, the demagnetizing field was not included ($H_D=0$) in Simon effect $S(H)$. Note that VelCrys gives the same value for the cases with $\boldsymbol{H}\parallel\boldsymbol{k}$ and $\boldsymbol{H}\perp\boldsymbol{k}$, failing to reproduce the small observed shift $\delta$ between both cases. In the case of transverse waves (Fig.\ref{fig:CoPt_trans}), this discrepancy is  due to the fact that Rinaldi and Turilli work\cite{rinaldi1985} is based on infinitesimal strain theory, while this shift arises from the rotational-magnetostrictive effect $(\delta=\delta R=3[\lambda_{100}-\lambda_{111}]=0.0008)$ based on the finite strain theory \cite{Rouchy1979,ROUCHY198069,booktremolet,nieves_sound2022}. Similarly, in the case of longitudinal wave (Fig.\ref{fig:CoPt_long}), this discrepancy is due to the lack of higher order corrections coming from morphic coefficients ($\delta=\delta G=\vert m_1^{\gamma,2}\vert/C_{11}=0.0005$) \cite{nieves_sound2022,booktremolet}. The low field regime below magnetic saturation is influenced by $\Delta s$-effect due to  domain walls displacement\cite{ROUCHY198069}, and cannot be accounted by Simon effect described by function $S(H)$ in Table \ref{table:theory}, as well as VelCrys, since they assume a magnetic saturated state\cite{Simon+1958+84+89,rinaldi1985}. In Fig.\ref{fig:CoPt_3d}, we show the landscape of the fractional change in group velocity for the 3 types of waves (qP, qS1 and qS2) as a function of the ray direction (phase velocity direction) with applied magnetic field $\mu_0\boldsymbol{H}=(0,0,5)$ T generated with VelCrys. We see that for the qP wave we have $(v-v_0)/v_0=0$ at $\boldsymbol{k}\parallel[001]$ consistent with the Simon effect ($S(H)=0$), see Table \ref{table:theory}. The largest value is observed in the qS1 wave about $(v-v_0)/v_0=-450\times10^{-6}$. Note that the values shown in Fig.\ref{fig:CoPt_3d} does not include the $\Delta s$-effect. Theoretical expression for other high symmetry ray and field directions were derived by Rouchy \textit{et al.}\cite{Rouchy1979,ROUCHY198069}, which can be also studied with VelCrys.

\begin{figure}[h!]
\centering
\includegraphics[width=0.8\columnwidth ,angle=0]{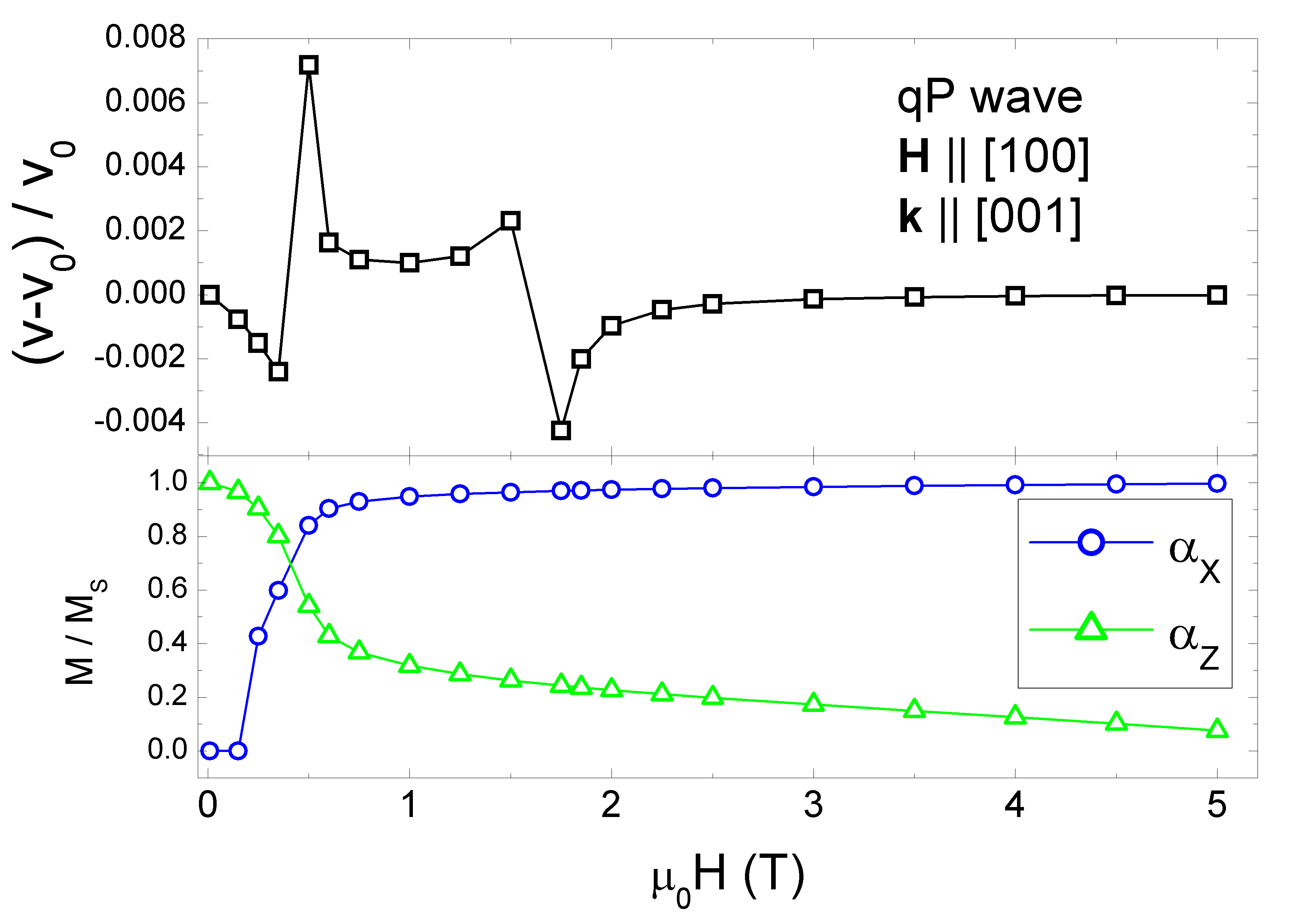}
\caption{(Top) Fractional change in group velocity $(v-v_0)/v_0$ for the qP wave with ray direction $\boldsymbol{k}\parallel[001]$ as a function of the applied magnetic field $\boldsymbol{H}\parallel[100]$  for hcp Co generated with VelCrys. (Bottom) The $x$ and $z$ components of direction cosines of equilibrium magnetization direction as function of the applied magnetic field.}
\label{fig:hcp_Co}
\end{figure}

\begin{figure}[h!]
\centering
\includegraphics[width=\columnwidth ,angle=0]{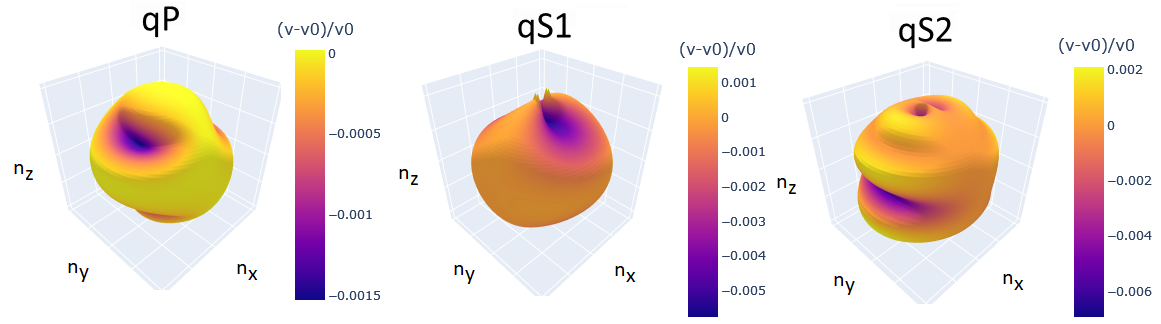}
\caption{Landscape of the fractional change in group velocity $(v-v_0)/v_0$ for the 3 types of waves (qP, qS1 and qS2) as a function of the ray direction (phase velocity direction) with applied magnetic field $\mu_0\boldsymbol{H}=(5,0,0)$ T for hcp Co generated with VelCrys. }
\label{fig:hcpCo_3d}
\end{figure}
\subsection{hcp Co}
\label{section:hcpCo}

Most of the Co produced is used for special alloys \cite{DEHAINE2021106656}. A relatively large percentage of the world’s production goes into magnetic alloys such as the Alnicos for permanent magnets. Sizable quantities are utilized for alloys that retain their properties at high temperatures and superalloys that are used near their melting points (where steels would become too soft). Co is also employed for hard-facing alloys, tool steels, low-expansion alloys (for glass-to-metal seals), and constant-modulus (elastic) alloys (for precision hairsprings). In this last example, we study magnetic corrections to group velocity for Co with hexagonal symmetry (hcp Co). We point out that magnetic field effect on group velocity is still rather unexplored in hexagonal crystals \cite{rinaldi1985,Danilevich_2018}. The used parameters in VelCrys for this material are given in Table \ref{table:data_properties_hcpCo}. 
\begin{table*}[h]
\caption{Experimental elastic constants ($C_{ij}$), anisotropic magnetoelastic constants ($b$), magnetocrystalline anisotropy constants ($K$), saturation magnetization ($M_s$) and mass density ($\rho$) for hcp Co \cite{Handley,maelas_publication2021}.}
\label{table:data_properties_hcpCo}
\centering
\resizebox{\textwidth}{!}{
\begin{tabular}{cc|cc|cc|c|c}
\toprule
 $C_{ij}$
&  (GPa)  &
$b$	& (MPa)  
& $K$ & (kJ/m$^3$) &  	 $\mu_0M_s$(T)  &  $\rho$(kg/m$^3$)\\

\midrule
 $C_{11}$  & 307  & $b_{21}$ &  -31.9 &$K_{1}$ & 410  &   1.76 &	8900 \\
                   $C_{12}$ & 165 & $b_{22}$ &  25.5 & $K_{2}$ & 150    &    \\
			$C_{13}$  & 103 & $b_3$ & -8.1 & & & &    \\
   $C_{33}$  & 358 & $b_4$ & 42.9 & & & &    \\
   $C_{44}$  & 75 & & & & & &    \\

\bottomrule
\end{tabular}
}
\end{table*}

In Fig.\ref{fig:hcp_Co}, we computed the fractional change in group velocity for the qP wave with ray direction $\boldsymbol{k}\parallel[001]$ as a function of the applied magnetic field $\boldsymbol{H}\parallel[100]$. At high fields ($\mu_0 H>2$T), we observe a field dependence $\sim 1/H$ consistent with Simon effect. In the low field regime ($\mu_0 H<2$T), a non-monotonic behaviour is observed probably due to an interplay between magnetic anisotropy, applied magnetic field, and rotation of magnetization. Similar peaks have been observed in Co-Pt alloys\cite{ROUCHY198069}. In Fig.\ref{fig:hcpCo_3d}, we present the landscape of the fractional change in group velocity for the 3 types of waves (qP, qS1 and qS2) as a function of the ray direction (phase velocity direction) with applied magnetic field $\mu_0\boldsymbol{H}=(5,0,0)$ T generated with VelCrys. Largest values are found in the qS2 wave ($\Delta v/v_0\sim -0.006$) along the direction of the magnetic field $\boldsymbol{k}\parallel\boldsymbol{H}\parallel[100]$. More theoretical analysis and experimental measurements in hexagonal crystals would be highly desirable to have a more deep understanding of these results \cite{Danilevich_2018}. Note that the same limitations of the implemented methodology as in the case of cubic crystals (lack of higher order corrections and $\Delta s$-effect) are also expected for the hexagonal symmetry.

\section{Conclusion}
\label{section:conclusion}

In summary, we showed that the combination of Wang \textit{et al.}\cite{wang2023} analytical solution of group velocity for a general elastic tensor with Rinaldi and Turilli\cite{rinaldi1985} effective magnetic corrections could provide a convenient approach to easily compute the group velocity and fractional change in group velocity due to magnetic effects. Such strategy has been implemented in the open source program VelCrys, facilitating the calculation of these quantities. VelCrys also includes visualization features to generate 3D plots of the landscape of these quantities as a function of the ray direction.

We have validated the program by recovering Wang \textit{et al.} results of group velocity for the dry sandstone \cite{wang2023}. The magnetic corrections\cite{rinaldi1985} were tested on cubic disordered Co-Pt alloy showing good agreement with theory and experiment after adding the $\Delta$s-effect\cite{ROUCHY198069}. We also applied this method to hcp Co finding a field dependence consistent with Simon effect\cite{Simon+1958+84+89}, as well as complex landscapes of fractional change in group velocity that have yet to be fully explained. Current version of the program has some limitations. For example, magnetic corrections to the elastic tensor has been implemented only for cubic I and hexagonal I symmetries. Additionally, these corrections does not include higher order terms associated to morphic coefficients and rotational-magnetostrictive effect, as well as $\Delta s$-effect related to domain walls displacement. Hopefully, some of these limitations might be overcome in future releases of the program. 

Potential applications of VelCrys could be in the research field of geophysics\cite{wang2023,REGAN1984227,CARA1987246,thomsen,Schimmel}, communication technology\cite{Yang_2022}, biosensors\cite{HUANG2021100041}, acoustic spin pumping \cite{Uchida2011}, magnetization switching induced by sound waves \cite{Camara,Kovalenko,Thevenard2013,Thevenard2016,Vlasov,Stupakiewicz2021,Weiyang,strungaru2023route}, and  picosecond ultrasonics \cite{MATTERN2023100503}. VelCrys could also be useful for further post-processing of elastic tensor and magnetoelastic constants obtained from automated schemes \cite{elate,deJong2015,AELAS,Maelas,maelasviewer}. Moreover, its web-based interface might be exploited in  databases of material science \cite{Mat_Proj_1,CURTAROLO2012218}.

\section*{Declaration of competing interest}

The authors declare that they have no known competing financial interests or personal relationships that could have appeared to influence the work reported in this paper.

\section*{Data availability}

Source files of VelCrys tool are available on the
GitHub repository \cite{VelCrys}.

\section*{Acknowledgement}

This work was supported by projects e-INFRA CZ (ID:90254)" and QM4ST (CZ.02.01.01/00/22\_008/0004572) by The Ministry of Education, Youth and Sports of the Czech Republic and also by Czech Science Foundation of the Czech Republic by grant No. 22-35410K.  P. N. acknowledges support by grant MU-23-BG22/00168 funded by The Ministry of Universities of Spain.
A.F. acknowledges funding from the Spanish Ministry of Science and
Innovation (grants nos.  PID2022-139230NB-I00, and TED2021-132074B-C32) the Diputación Foral de Gipuzkoa (Project No. 2023-CIEN-000077-01). Research conducted in the scope of the Transnational Common Laboratory (LTC) Aquitaine-Euskadi Network in Green Concrete and Cement-based Materials. R. I. acknowledges financial support from the project MAGNES funded by the Principality of Asturias Government (grant no. AYUD/2021/51822) and from and the project RADIAFUS V, funded by the Spanish Ministry of Science and Innovation (grant no. PID2019-105325RB-C32).

\bibliographystyle{elsarticle-num}
\bibliography{mybibfile.bib}







\end{document}